\documentclass[aps,onecolumn,showpacs]{revtex4}
\usepackage{graphicx}
\usepackage{amsmath}
\usepackage{slashed}
\usepackage{bbold}
\include{amssym}
\def\PSfig#1#2{\centerline{\scalebox{#1}{\includegraphics{#2}}}}

\def\ca{\c{c}\~{a}}

\def\ii{\'{\i}}

\newcommand{\dt}{{\mathrm{det}}}

\newcommand{\ud}{\mathrm{d}}
\newcommand{\st}{{\mathrm{st}}}

\newcommand{\eff}{\mathrm{eff}}

\begin{document}

\title{\bf The phase diagram for the Nambu--Jona-Lasinio model with 
           't Hooft and eight-quark interactions} 

\author{B. Hiller, J. Moreira, A. A. Osipov\footnote{On leave from 
        Dzhelepov Laboratory of Nuclear Problems, 
        Joint Institute for Nuclear Research, 
        141980 Dubna, Moscow Region, Russia}, A. H. Blin}
%  Email address: osipov@nu.jinr.ru 
%  Email address: brigitte@teor.fis.uc.pt 
%  Email address: jmoreira@teor.fis.uc.pt
%  Email address: alex@teor.fis.uc.pt
\affiliation{Centro de F\'{\i}sica Computacional, Departamento de
             F\'{\i}sica da Universidade de Coimbra, 3004-516 Coimbra, 
             Portugal}

\begin{abstract}
It is shown that the endpoint of the f\mbox{}irst order transition line 
which merges into a crossover regime in the phase diagram of the 
Nambu--Jona-Lasinio model, extended to include the six-quark 't Hooft 
and eight-quark interaction Lagrangians, is pushed towards vanishing 
chemical potential and higher temperatures with increasing strength of 
the OZI-violating eight-quark interactions. We clarify the connection
between the location of the endpoint in the phase diagram and the mechanism 
of chiral symmetry breaking at the quark level. We show how the $8q$ 
interactions af\mbox{}fect the number of ef\mbox{}fective quark degrees of 
freedom. We are able to obtain the correct asymptotics for this number 
at large temperatures by using the Pauli-Villars regularization. 
\end{abstract}

\maketitle

%%%%%%%%%%%%%%%%%%%%%%%%%%%%%%%%%%%%%%%%%%%%%%%%%%%%%%%%%%%%%%%%%%%%%%%%%%%%
%                        main text
%%%%%%%%%%%%%%%%%%%%%%%%%%%%%%%%%%%%%%%%%%%%%%%%%%%%%%%%%%%%%%%%%%%%%%%%%%%%

\section{Introduction}
The last two decades have witnessed great ef\mbox{}forts towards the 
understanding of the QCD phase diagram, in terms of ef\mbox{}fective low energy
theories paralleled by QCD lattice calculations, see e.g. the recent reviews 
\cite{Wilczek:2001}-\cite{Fukushima:2008}, or the paper \cite{Weise:2006}. The 
domain of small to moderate baryonic chemical potential $0<\mu <400$ MeV and 
temperatures $0<T < 200$ MeV, is of specif\mbox{}ic relevance for relativistic 
heavy ion collisions. Of fundamental importance in the study of the phase 
diagram are chiral symmetry and conf\mbox{}inement, however the f\mbox{}inite 
size ($m_u, m_d, m_s\neq 0$) as well as the dif\mbox{}ference in the bare quark
masses ($m_u\neq m_d\neq m_s$) pose major problems both from the calculational 
point of view and in the implications due to deviations from the ideal 
situations, where the chiral condensate and the Polyakov loop are known to be 
the appropriate order parameters to characterize the phase state of the 
quark-gluon system.

The present study focuses on the chiral symmetry breaking aspects related to 
non-zero current quark mass values. Our arguments will be based on the 
successful model of Nambu--Jona-Lasinio (NJL) \cite{Nambu:1961}, combined with 
the $U_A(1)$ breaking $2N_f$ f\mbox{}lavor determinant of 't Hooft 
\cite{Hooft:1976}-\cite{Reinhardt:1988} (NJLH). Moreover the most general 
$U(3)_L\otimes U(3)_R$ chiral invariant non derivative eight-quark ($8q$) 
interactions \cite{Osipov:2005b} are included. These terms were proven to 
render the ef\mbox{}fective potential of the NJLH model globally stable, and 
their ef\mbox{}fect has been thoroughly studied in low energy characteristics 
of pseudoscalar and scalar mesons \cite{Osipov:2006a}, at f\mbox{}inite 
temperature \cite{Osipov:2007b,Osipov:2008} and in presence of a constant 
magnetic f\mbox{}ield \cite{Osipov:2007c}. These studies have lead at instances
to sizeable and unforeseen ef\mbox{}fects. Of particular importance for the 
present work is that the strength of the $8q$ coupling is strongly correlated 
with the temperature and slope at which the crossover occurs and that it can 
be regulated together with the four-quark coupling, leaving the meson spectra 
at $T=0$ unaf\mbox{}fected (with exception of the scalar $\sigma$ meson mass 
which decreases with increasing $8q$ coupling) \cite{Osipov:2008}. As a result 
the symmetry breaking for large $8q$ couplings is induced by the $6q$ 't Hooft 
coupling strength, as opposed to the case with small $8q$ coupling, where the 
dynamical breaking of symmetry is controlled by the $4q$ coupling strength 
\cite{Osipov:2006a}. We would like to comment on a natural question which 
arises here, namely, can higher order many-quark interactions be also 
important? With regard to this, explicit argumets of A. A. Andrianov and
V. A. Andrianov are known \cite{Andrianov:1993}, which show that the structure 
of the QCD-motivated models at low energies with ef\mbox{}fective multi-fermion 
interactions and a f\mbox{}inite cut-of\mbox{}f in the chiral symmetry-breaking
regime should contain only the vertices with four, six and eight-fermion 
interactions in four dimensions. This result explains partly the approximation 
used in our work. 

Thus, in this paper we give a quantitative account of local multi-fermion 
forces on the phase diagram in the $(T,\mu )$ plane, by comparing the results 
for two sets of parameters, in the small and large coupling regimes of the 
$8q$ strengths, corresponding to the two above mentioned alternative 
mechanisms of chiral symmetry breaking.

Throughout the paper we work for simplicity in the isospin limit $m_u=m_d
\ne m_s$,  breaking explicitly the chiral $SU(3)_L\otimes SU(3)_R$ 
symmetry to the $SU(2)_I\otimes U(1)_Y$ (isospin-hypercharge) subgroup, and 
take the same baryonic chemical potential $\mu$ for all quark species. 
Generalizations to take into account the nonzero isospin chemical potential 
can be implemented as for instace in \cite{Klimenko:2006}.    

\section{Effective Lagrangian}
The explicit form of the multi-quark Lagrangian considered is presented in 
\cite{Osipov:2005b,Osipov:2006a} 
\begin{equation}
\label{efflag}
  {\cal L}_\eff =\bar{q}(i\gamma^\mu\partial_\mu - m)q
          +{\cal L}_{4q} + {\cal L}_{6q}
          +{\cal L}_{8q}^{(1)}+{\cal L}_{8q}^{(2)}.
\end{equation}
Quark f\mbox{}ields $q$ have color $(N_c=3)$ and f\mbox{}lavor $(N_f=3)$
indices which are suppressed, $\mu =0,1,2,3$. Here
\begin{eqnarray}
\label{L4q}
   {\cal L}_{4q} &\!\! =\!\! & \frac{G}{2}\left[(\bar{q}\lambda_aq)^2+
                      (\bar{q}i\gamma_5\lambda_aq)^2\right], \\
\label{Ldet}
   {\cal L}_{6q} &\!\! =\!\! & \kappa (\mbox{det}\ \bar{q}P_Lq
                      + \mbox{det}\ \bar{q}P_Rq), \\  
   {\cal L}_{8q}^{(1)}&\!\! = \!\! & 
   8g_1\left[ (\bar q_iP_Rq_m)(\bar q_mP_Lq_i) \right]^2, \\ 
   {\cal L}_{8q}^{(2)}&\!\! = \!\!& 
   16 g_2\left[ (\bar q_iP_Rq_m)(\bar q_mP_Lq_j) 
   (\bar q_jP_Rq_k)(\bar q_kP_Lq_i) \right]. 
\end{eqnarray}
The matrices acting in f\mbox{}lavor space, $\lambda_a,\ a=0,1,\ldots ,8,$ 
are normalized such that $\mbox{tr} (\lambda_a \lambda_b )=2\delta_{ab}$; 
$\lambda_0=\sqrt{\frac{2}{3}}\, 1$, and $\lambda_k,\ k=1,2,\ldots ,8$ are the 
standard $SU(3)$ Gell-Mann matrices; $P_{L,R}=(1\mp\gamma_5)/2$ are chiral 
projectors and the determinant is over f\mbox{}lavor indices. The large $N_c$ 
behaviour of the model is ref\mbox{}lected in the dimensionful coupling 
constants, $[G]=M^{-2},\, [\kappa ]=M^{-5},\, [g_1]=[g_2]=M^{-8}$, which count 
as $G\sim 1/N_c$, $\kappa\sim 1/N_c^{N_f}$, and $g_1,g_2\sim 1/N_c^4$ or less. 
As a result the NJL interaction (\ref{L4q}) dominates over ${\cal L}_{6q}$ at 
large $N_c$, as one would expect, because Zweig's rule is exact at $N_c = 
\infty$. Let us note that the $8q$-interaction ${\cal L}_{8q}^{(1)}$ breaks 
Zweig's rule as well. 

Since the coupling constants $G, \kappa, g_1, g_2$ are dimensionful, the model 
is not renormalizable. We use the cut-of\mbox{}f $\Lambda$ to render quark 
loops f\mbox{}inite. The global chiral $SU(3)_L\times SU(3)_R$ symmetry of the 
Lagrangian (\ref{efflag}) at $m=0$ is spontaneously broken to the $SU(3)$ 
group, showing the dynamical instability of the fully symmetric solutions of 
the theory. In addition, the current quark mass $m$, being a diagonal matrix 
in f\mbox{}lavor space with elements $\mbox{diag} (m_u, m_d, m_s)$, explicitly 
breaks this symmetry down, retaining only the reduced $SU(2)_I\times U(1)_Y$ 
symmetries of isospin and hypercharge conservation, if $m_u = m_d \neq m_s$. 

The model has been bosonized in the framework of functional integrals in the 
stationary phase approximation leading to the following ef\mbox{}fective 
mesonic Lagrangian ${\cal L}_{{\rm bos}}$ at $T=\mu=0$
\begin{eqnarray}
\label{bos}
    &&{\cal L}_\eff\to {\cal L}_{{\rm bos}}={\cal L}_{\st}+{\cal L}_{{\rm ql}}, 
    \qquad
    {\cal L}_{\st}=h_a\sigma_a+\frac{1}{2} h_{ab}^{(1)}
                  \sigma_a\sigma_b+\frac{1}{2} h_{ab}^{(2)}
                  \phi_a\phi_b +{\cal O}({\mbox{f\mbox{}ield}}^3),
                  \nonumber\\
    &&W_{{\rm ql}}(\sigma,\phi )=\frac{1}{2}\mbox{ln}|\dt 
                 D^{\dagger}_E D_E|=-\!\int\!
                 \frac{\ud^4 x_E}{32 \pi^2}
                 \sum_{i=0}^\infty I_{i-1}\mbox{tr}(b_i)=\int\!\ud^4 x_E
                 {\cal L}_{{\rm ql}},
                 \nonumber\\
    &&b_0 =1,\hspace{0.2cm} b_1=-Y,
                 \hspace{0.2cm} b_2=\frac{Y^2}{2} 
                 +\frac{\Delta_{us}}{\sqrt{3}}\lambda_8 
                 Y, \hspace{0.2cm} \ldots , 
                 \nonumber\\
    &&Y=i\gamma_{\alpha}(\partial_{\alpha}\sigma 
                 +i\gamma_5\partial_{\alpha}\phi )
                 +\sigma^2+\{{\cal M},\sigma\}+\phi^2 
                 +i\gamma_5[\sigma+{\cal M},\phi ]
\end{eqnarray}
written in terms of the scalar, $\sigma=\lambda_a\sigma_a$, and pseudoscalar, 
$\phi=\lambda_a\phi_a$, nonet valued quantum f\mbox{}ields. The result of the 
stationary phase integration at leading order, ${\cal L}_\st$, is shown here as
a series in growing powers of $\sigma$ and $\phi$. The result of the remaining 
Gaussian integration over the quark f\mbox{}ields is given by $W_{\rm ql}$. Here
the Laplacian in euclidean space-time ${D^{\dagger}_E D}_E={\cal M}^2-
\partial_\alpha^2+Y$ is associated with the euclidean Dirac operator $D_E=i
\gamma_\alpha\partial_\alpha -{\cal M}-\sigma -i\gamma_5\phi$ (the $\gamma_\alpha,
\,\alpha =1,2,3,4$ are antihermitian and obey $\{\gamma_\alpha,\gamma_\beta\}=-2
\delta_{\alpha\beta}$); ${\cal M}=\mbox{diag}(M_u,M_d,M_s)$ is the constituent 
quark mass matrix (to explore the properties of the spontaneously broken theory,
we def\mbox{}ine quantum f\mbox{}ields $\sigma_a, \phi_a$ as having vanishing 
vacuum expectation values in the asymmetric phase).

The expression for the one-quark-loop action $W_{\rm ql}$ has been obtained 
by using a modif\mbox{}ied inverse mass expansion of the heat kernel 
associated to the given Laplacian \cite{Osipov:2001}. The procedure takes into 
account the dif\mbox{}ferences $\Delta_{us}=M_u^2-M_s^2$ in the nonstrange and 
strange constituent quark masses in a chiral invariant way at each order of 
the expansion, $b_i$ being the generalized Seeley--DeWitt 
coef\mbox{}f\mbox{}icients of the new series. This modif\mbox{}ication 
distinguishes our calculation from the one made in \cite{Ebert:1986}. In fact 
we consider the series up to and including the order $b_2$ that corresponds to 
the f\mbox{}irst nontrivial step in the expansion of the induced 
ef\mbox{}fective hadron Lagrangian at long distances. At this stage meson 
f\mbox{}ields obtain their kinetic terms, but are still considered to be 
elementary objects. The information about their quark-antiquark origin enters 
only through the coef\mbox{}f\mbox{}icients such as the average
\begin{equation}
\label{I-i}
   I_i=\frac{1}{3}\left[J_i(M_u^2)+J_i(M_d^2)+J_i(M_s^2)\right]
\end{equation}
over the 1-loop euclidean momentum integrals $J_i$ with $i+1$ vertices 
($i=0,1,\ldots$) 
\begin{equation} 
\label{Ji}
    J_i(M^2)=16\pi^2\Gamma (i+1)\!\int %_{{\mathrm R}}
             \frac{\ud^4p_E}{(2\pi)^4}\,\hat\rho_\Lambda 
             \frac{1}{(p_E^2+M^2)^{i+1}}. 
\end{equation}  
For the explicit evaluation of $J_i(M^2)$ we use the Pauli--Villars 
regularization method with two subtractions in the integrand. The 
procedure is fully def\mbox{}ined by the insertion of the particular 
operator
\begin{equation}
\label{reg-2}
    \hat\rho_\Lambda =1-\left(1-\Lambda^2\frac{\partial}{\partial M^2}
    \right) \exp\left(\Lambda^2\frac{\partial}{\partial M^2}\right).       
\end{equation}     
Here the covariant cut-of\mbox{}f $\Lambda$ is a free dimensionful parameter 
which characterizes the scale of the chiral symmetry breaking in the 
ef\mbox{}fective model considered. To the order of the heat kernel series 
truncated, only the integrals $J_0,J_1$ are needed. These are quadratic and 
logarithmic divergent respectively with $\Lambda\to\infty$, all other $J_i$ 
are f\mbox{}inite. Note that the recurrence relation
\begin{equation}
\label{rec}
    J_{i+1}(M^2)=-\frac{\partial}{\partial M^2} J_i(M^2)
\end{equation}
is fulf\mbox{}illed. If $J_i$ is known for one value of $i$, then the function 
may be computed for other values of $i$ by successive applications of the 
relation.  

In ${\cal L}_{\st}$ the $h_a$ are determined via the stationary phase conditions.
These conditions and the pattern of explicit symmetry breaking show that in 
general $h_a$ can have only three non-zero components at most with indices 
$a=0,3,8$, i.e. $h_a\lambda_a=\mbox{diag} (h_u,h_d,h_s)$, which can be found
from a system of three independent equations  
\begin{equation}
\label{SPA}
   \left\{ \begin{array}{l}
\vspace{0.2cm}   
   Gh_u + \Delta_u +\displaystyle\frac{\kappa}{16}\ h_dh_s
   +\displaystyle\frac{g_1}{4}\ h_u(h_u^2+h_d^2+h_s^2)
   +\displaystyle\frac{g_2}{2}\ h_u^3=0, \\
\vspace{0.2cm}   
   Gh_d + \Delta_d +\displaystyle\frac{\kappa}{16}\ h_uh_s
   +\displaystyle\frac{g_1}{4}\ h_d(h_u^2+h_d^2+h_s^2)
   +\displaystyle\frac{g_2}{2}\ h_d^3=0, \\
\vspace{0.2cm}   
   Gh_s + \Delta_s +\displaystyle\frac{\kappa}{16}\ h_uh_d
   +\displaystyle\frac{g_1}{4}\ h_s(h_u^2+h_d^2+h_s^2)
   +\displaystyle\frac{g_2}{2}\ h_s^3=0. 
   \end{array} \right.
\end{equation}
Here $\Delta_f=M_f-m_f$, $f=u,d,s$. The matrix valued constants of higher 
order, like for instance $h_{ab}^{(1,2)}$ in ${\cal L}_\st$, are uniquely 
determined once the $h_f$ are known \cite{Osipov:2004a,Osipov:2006a}. The 
stability of the ef\mbox{}fective potential is guaranteed if the system 
(\ref{SPA}) has only one real solution. For that the couplings must 
fulf\mbox{}ill the inequalities \cite{Osipov:2005b}: $g_1>0,\, g_1+3g_2>0,\, 
Gg_1>(\kappa/16)^2$.

\vspace{1.0cm}

\noindent
{\small TABLE I. 
Parameters of the model: $m_u=m_d,\, m_s$ (MeV), $G$ (GeV$^{-2}$), $\Lambda$ 
(MeV), $\kappa$ (GeV$^{-5}$), $g_1,\, g_2$ (GeV$^{-8}$). We also show the 
corresponding values of constituent quark masses $M_u=M_d$ and $M_s$ (MeV).
}

\noindent
\begin{tabular}{lccccccccccc}
\hline\hline 
&$\ \ \ \ \ \ \ \ \ \ \ \ \ m_u$  
&$\ \ \ \ \ \ \ \ \ \ \ m_s$  
&$\ \ \ \ \ \ \ \ \ \ M_u$  
&$\ \ \ \ \ \ \ \ \ \ M_s$  
&$\ \ \ \ \ \ \ \ \ \ \Lambda$ 
&$\ \ \ \ \ \ \ \ \ G$ 
&$\ \ \ \ \ \ \ \ -\kappa$ 
&$\ \ \ \ \ \ \ \ \ g_1$
&$\ \ \ \ \ \ \ \ \ -g_2\ \ $   
\\ \hline
\ a 
&\ \ \ \ \ \ \ \ \ \ \ \ \ 5.9
&\ \ \ \ \ \ \ \ \ \ 186
&\ \ \ \ \ \ \ \ \ \ 359
&\ \ \ \ \ \ \ \ \ \ 554
&\ \ \ \ \ \ \ \ \ \ 851
&\ \ \ \ \ \ \ \ 10.92
&\ \ \ \ \ \ \ \ \ 1001
&$\ \ \ \ \ \ \ \ \ 1000^*$
&$\ \ \ \ \ \ \ \ 47\ \ $
\\
\ b 
&\ \ \ \ \ \ \ \ \ \ \ \ \ 5.9
&\ \ \ \ \ \ \ \ \ \ 186
&\ \ \ \ \ \ \ \ \ \ 359
&\ \ \ \ \ \ \ \ \ \ 554
&\ \ \ \ \ \ \ \ \ \ 851
&\ \ \ \ \ \ \ \ \ 7.03
&\ \ \ \ \ \ \ \ \ 1001
&$\ \ \ \ \ \ \ \ \ 8000^*$
&$\ \ \ \ \ \ \ \ 47\ \ $
\\ 
\hline\hline
\end{tabular}

\vspace{1.0cm}
  
\noindent
{\small TABLE II. 
The masses, weak decay constants of light pseudoscalars (in MeV), the 
singlet-octet mixing angle $\theta_p$ (in degrees), and the quark condensates 
$\langle\bar uu\rangle =\langle\bar dd\rangle, \langle\bar ss\rangle$ 
expressed as usual by positive combinations in MeV. 
}

\noindent
\begin{tabular}{lcccccccccc}
\hline\hline
&$\ \ \ \ \ \ \ \ \ \ \ \ \ m_\pi$
&$\ \ \ \ \ \ \ \ \ m_K$
&$\ \ \ \ \ \ \ \ m_\eta$
&$\ \ \ \ \ \ \ \ \ m_{\eta'}$
&$\ \ \ \ \ \ \ \ \ f_\pi$
&$\ \ \ \ \ \ \ \ \ f_K$
&$\ \ \ \ \ \ \ \ \ \theta_p$
&$\ \ \ \ \ \ \ -\langle\bar uu\rangle^{\frac{1}{3}}$ 
&$\ \ \ \ \ \ \ -\langle\bar ss\rangle^{\frac{1}{3}}$
\\ 
\hline
\ a 
&$\ \ \ \ \ \ \ \ \ \ \ \ \ 138^*$
&$\ \ \ \ \ \ \ \ \ 494^*$
&$\ \ \ \ \ \ \ \ 477$
&$\ \ \ \ \ \ \ \ \ 958^*$
&$\ \ \ \ \ \ \ \ \ 92^*$
&$\ \ \ \ \ \ \ \ \ 117^*$
&$\ \ \ \ \ \ \ \ -14$
&$\ \ \ \ \ \ \ 235$
&$\ \ \ \ \ \ \ 187\ \ $
\\ 
\ b 
&$\ \ \ \ \ \ \ \ \ \ \ \ \ 138^*$
&$\ \ \ \ \ \ \ \ \ 494^*$
&$\ \ \ \ \ \ \ \ 477$
&$\ \ \ \ \ \ \ \ \ 958^*$
&$\ \ \ \ \ \ \ \ \ 92^*$
&$\ \ \ \ \ \ \ \ \ 117^*$
&$\ \ \ \ \ \ \ \ -14$
&$\ \ \ \ \ \ \ 235$
&$\ \ \ \ \ \ \ 187\ \ $ 
\\ 
\hline\hline
\end{tabular}

\vspace{1.0cm}

\noindent
{\small TABLE III. 
The masses of the scalar nonet (in MeV), and the corresponding singlet-octet 
mixing angle $\theta_s$ (in degrees). 
}

\noindent
\begin{tabular}{lccccc}
\hline\hline
&$\ \ \ \ \ \ \ \ \ \ \ \ \ \ \ \ \ \ \ \ m_{a_0(980)}$
&$\ \ \ \ \ \ \ \ \ \ \ \ \ \ \ \ \ \ m_{K_0^*(800)}$
&$\ \ \ \ \ \ \ \ \ \ \ \ \ \ \ \ \ m_{f_0(600)}$ 
&$\ \ \ \ \ \ \ \ \ \ \ \ \ \ \ \ \ m_{f_0(1370)}$
&$\ \ \ \ \ \ \ \ \ \ \ \ \ \ \ \theta_s\ $
\\ 
\hline
\ a 
&$\ \ \ \ \ \ \ \ \ \ \ \ \ \ \ \ \ \ \ \ 980^*$ 
&$\ \ \ \ \ \ \ \ \ \ \ \ \ \ \ \ \ 1201$
&$\ \ \ \ \ \ \ \ \ \ \ \ \ \ \ \ 691$
&$\ \ \ \ \ \ \ \ \ \ \ \ \ \ \ \ 1368$
&$\ \ \ \ \ \ \ \ \ \ \ \ \ \ \ \ \ \ 23\ \ \ \ $ 
\\ 
\ b 
&$\ \ \ \ \ \ \ \ \ \ \ \ \ \ \ \ \ \ \ \ 980^*$ 
&$\ \ \ \ \ \ \ \ \ \ \ \ \ \ \ \ \ 1201$
&$\ \ \ \ \ \ \ \ \ \ \ \ \ \ \ \ 463$
&$\ \ \ \ \ \ \ \ \ \ \ \ \ \ \ \ 1350$
&$\ \ \ \ \ \ \ \ \ \ \ \ \ \ \ \ \ \ 19\ \ \ \ $ 
\\
%f & 977  & 1201  & 730  & 1372 & 25 \\
\hline\hline
\end{tabular}
\vspace{1.0cm}

In this paper we use the two parameter sets of Table I, which dif\mbox{}fer 
only in the choice of the $4q$ coupling $G$ and the $8q$ strength $g_1$. Set 
(b) is the same as in \cite{Osipov:2008} (there was a misprint in the value 
for the constituent strange quark mass, which we corrected). Tables II-III 
display the numerical f\mbox{}its at $T=\mu=0$ (input is denoted by a *). The 
only dif\mbox{}ference in the observables of the two sets occurs in the 
singlet-octet f\mbox{}lavor mixing channel of the scalars, mainly in the 
$\sigma$-meson (i.e. $f_0(600)$) mass. The model parameters are kept unchanged 
in the calculation of the $T$ and $\mu$ dependent solutions of the gap 
equations (see next sections). 

It is worthwhile to stress that there is an essential dif\mbox{}ference 
between the two alternative ground states chosen here as the 
conf\mbox{}igurations on top of which the $T\neq 0$ and $\mu\neq 0$ 
ef\mbox{}fects are studied: Case (a) corresponds to the standard picture of 
the NJL hadronic vacuum. In this picture chiral symmetry is spontaneously 
broken at $T=\mu =0$ when $G>G_{crit}$. Case (b) corresponds to a new 
alternative, related to the pattern where $G<G_{crit}$. In this case chiral 
symmetry can be broken only due to the six-quark interactions, when $|\kappa |$ 
exceeds some critical value (the $8q$-interactions could in principle also 
induce symmetry breaking, however the mass spectra are then not well 
reproduced). One can hardly distinguish between the two cases at $T=\mu=0$, 
the spectra of $0^{-+}$ and $0^{++}$ low-lying mesons do not show much 
dif\mbox{}ference: the model parameters are the same, except for the 
correlated $G$ and $g_1$ values. The larger value of $g_1$ in the case (b) is 
a signal of the increasing role played by the eight-quark OZI-violating 
interactions, but this does not af\mbox{}fect the value of the mixing angle 
$\theta_p$, and only slightly diminishes $\theta_s$. Such insensitivity follows
from the observation that the stationary phase equations (\ref{SPA}) and mass 
formulae of the light $0^{-+}$ and $0^{++}$ states \cite{Osipov:2006a} only 
depend on the couplings $G$ and $g_1$ through the linear combination $\xi=G+
g_1(h_u^2+h_d^2+h_s^2)/4$, except for the $00,\, 08$ and $88$ states inside 
the scalar nonet. However as soon as $T$ or $\mu$ are f\mbox{}inite, the $h_f$ 
start to change due to their intrinsic $T,\mu$ dependence, acquired through 
the coupling to the quark loop integrals in the gap equations, eqs. (\ref{gap})
below. The $T,\mu$ dependence of the combination $\xi$ above is steered by the 
strength $g_1$. This is the main reason why the $8q$-interactions may strongly 
af\mbox{}fect the thermodynamic observables, without changing the spectra at 
$T=\mu=0$. 
 
\section{Thermodynamic potential}
\subsection{Case of vanishing temperarure and chemical potential}
Before addressing the thermodynamical potential it is instructive to briefly 
discuss the ef\mbox{}fective potential of the model at $T=\mu=0$. Using 
standard techniques \cite{Osipov:2004a}, we obtain from the gap-equations 
\begin{equation}
\label{gap}
            h_f + \frac{N_c}{2\pi^2} M_f J_0(M_f^2)=0
\end{equation}
the ef\mbox{}fective potential $V(M_f)$ as a function of three independent 
variables $M_f=\{M_u,M_d,M_s\}$. If the parameters of the model are 
f\mbox{}ixed in such a way that eqs. (\ref{SPA}) have only one real solution, 
the ef\mbox{}fective potential is
\begin{eqnarray} 
\label{effpot1}
     V(M_f)&\!\! =\!\!&-\frac{1}{2}\int_0^{M_f}\!\!\sum_{f=u,d,s}h_f \ud M_f
           -\frac{N_c}{4\pi^2}\int_0^{M_f}\!\!\sum_{f=u,d,s}M_fJ_0(M_f^2)\ud M_f
           \nonumber \\
           &\!\! = \!\!&\frac{1}{16}\left.\left(4Gh_f^2+\kappa h_uh_dh_s 
           +\frac{3g_1}{2}\left(h_f^2\right)^2+3g_2h_f^4\right)\right|^{M_f}_0 
           +\frac{N_c}{8\pi^2}\sum_{f=u,d,s}\! J_{-1}(M_f^2),
\end{eqnarray}
where $h_f^2=h_u^2+h_d^2+h_s^2, \,\, h_f^4=h_u^4+h_d^4+h_s^4$, and we extend
def\mbox{}inition (\ref{I-i}) for index $i=-1$ with 
\begin{equation}
\label{J_1}
     J_{-1}(M^2)=-\int_0^{M^2}\!\! J_0(M^2)\ud M^2
     =-\frac{1}{2}\left(M^2J_0(M^2)+\Lambda^4\ln
     \left(1+\frac{M^2}{\Lambda^2}\right)\right).
\end{equation}
Here $J_0$ has the explicit form 
\begin{equation}
     J_0(M^2)=\Lambda^2 -M^2\ln\left(1+\frac{\Lambda^2}{M^2}\right)
\end{equation}
for the given choice of regulator. 

The f\mbox{}irst integral in (\ref{effpot1}) accounts for the leading order 
stationary phase contribution. The second integral describes the quark one-loop 
part. Since both integrands in (\ref{effpot1}) are exact 
dif\mbox{}ferentials, the line integrals depend only on the end points. The 
low limit of the integrals is adjusted so that $V(0)=0$ (to understand this,
it is enough to notice that the power-series expansion of $V(M_f)$ at small 
$M_f$ starts from $V(0)$; this term does not depend on $M_f$ and, therefore, 
does not af\mbox{}fect the physical content of the theory; so we simply 
subtract it, calculating the potential energy of the system with regard to 
the energy of the symmetric vacuum in the imaginary world of massless quarks).
   
\subsection{Case of finite temperature and chemical potential}
The extension to f\mbox{}inite $T$ and $\mu$ of the bosonized Lagrangian 
(\ref{bos}) is ef\mbox{}fected through the quark loop integrals $J_i$ (see 
eq.(\ref{Ji})). Due to the recurrence relation (\ref{rec}) it is 
suf\mbox{}f\mbox{}icient to get it just for one of them, $J_0$, by introducing 
the Matsubara frequencies, $\omega_n$, and the chemical potential, $\mu$, 
through the substitutions \cite{Kapusta}
\begin{equation}
\label{subst}
      \int\!\ud p_{0E}\to 2\pi T\!\!\!\sum_{n=-\infty}^\infty,
      \quad p_{0E}\to\omega_n-i\mu,\quad\omega_n= \pi T (2n+1).
\end{equation}
Inserting (\ref{subst}) into $J_0$ we obtain
\begin{eqnarray}
    J_0(M^2)&\!\!\to\!\! &J_0(M^2,T,\mu )=16\pi^2 T\int
    \frac{\ud^3\vec{p}_E}{(2\pi )^3}\hat{\rho}_\Lambda
    \sum_{n=-\infty}^{+\infty}\frac{1}{C_n-2i\mu\pi T(2n+1)}
    \nonumber\\
    C_n&\!\! =\!\!&E_p^2+\pi^2 T^2 \left(2n+1\right)^2 -\mu^2, \quad
    E_p = \sqrt{M^2+\vec{p}^{\, 2}_E}.
\end{eqnarray}
The sum over $n$ is evaluated to give
\begin{equation}
    \sum_{n=-\infty}^{+\infty}\frac{1}{C_n-i2\mu\omega_n}
    =\frac{1}{(2\pi T)^2}\sum^{+\infty}_{n=-\infty}\frac{1}{(n+a)(n+b)}
    =\frac{1}{4TE_p}\left(\tanh\frac{\mu +E_p}{2T}-\tanh \frac{\mu 
    -E_p}{2T}\right)
\end{equation}
where we use the abbreviations $a=\frac{1}{2}+\frac{i}{2\pi T}\left(E_p-\mu
\right)$ and $b=\frac{1}{2}-\frac{i}{2\pi T}\left(E_p+\mu\right)$.
Thus we have
\begin{equation}
\label{J0t}
     J_0(M^2,T,\mu )=4\int_0^\infty\ud |\vec{p}_E| |\vec{p}_E|^2
     \hat{\rho}_\Lambda\frac{1}{E_p}\left(1-n_{q}-n_{\overline{q}}\right)
     =J_0(M^2)- 4\int_0^\infty\ud |\vec{p}_E| |\vec{p}_E|^2
     \hat{\rho}_\Lambda\frac{n_{q}+n_{\overline{q}}}{E_p}
\end{equation}
where the quark, anti-quark occupation numbers are given by
\begin{equation}
\label{n-occup}
     n_{q}=\frac{1}{1+e^{\frac{E_p-\mu}{T}}}, \quad 
     n_{\overline{q}}=\frac{1}{1+e^{\frac{E_p+\mu}{T}}}.
\end{equation}
Notice that $J_0(M^2,0,0)=J_0(M^2)$, therefore the vacuum piece is well 
isolated from the matter part. The remaining integral containing the quark 
number occupation densities $n_q, n_{\bar q}$ is strictly f\mbox{}inite, the 
$\Lambda$ dependent terms being a remnant of the Pauli--Villars regularization 
scheme. At small $T$ and $M\neq 0$ we have
\begin{eqnarray}
\label{J0-asymp}
     \int_0^\infty\!\ud |\vec{p}_E| |\vec{p}_E|^2\frac{n_{q}}{E_p}
     &\!\!\simeq\!\!&
     \int_0^\infty\!\ud |\vec{p}_E| \frac{|\vec{p}_E|^2}{E_p}e^{-\frac{E_p}{T}}
     =T\sqrt{2TM}e^{-\frac{M}{T}}\int_0^\infty\!\ud x\sqrt{x}e^{-x}\sqrt{1
     +\frac{xT}{2M}} \nonumber \\
     &\!\! =\!\!&\sqrt{\frac{\pi M}{2}}T^{\frac{3}{2}}e^{-\frac{M}{T}}
     \left(1+\frac{3T}{8M}+\ldots \right).
\end{eqnarray}
The special feature of this integral is that it vanishes exponentially with 
$T\to 0$.

Now we are ready to evaluate the thermodynamical potential. Indeed, the 
gap-equations at f\mbox{}inite $T$ and $\mu$ are 
\begin{equation}
\label{gap-t}
            h_f + \frac{N_c}{2\pi^2} M_f J_0(M_f^2,T,\mu )=0.
\end{equation}
Consequently, 
\begin{eqnarray} 
\label{effpot-t}
     V(M_f,T,\mu )&\!\! =\!\!& -\frac{1}{2}\int_0^{M_f}\!\!\sum_{f=u,d,s}h_f\ud 
     M_f-\frac{N_c}{4\pi^2}\int_0^{M_f}\!\!\sum_{f=u,d,s}M_fJ_0(M_f^2,T,\mu )\ud 
     M_f + C(T,\mu )\nonumber \\
     &\!\! = \!\!&\frac{1}{16}\left.\left(4Gh_f^2+\kappa h_uh_dh_s 
     +\frac{3g_1}{2}\left(h_f^2\right)^2+3g_2h_f^4\right)\right|_0^{M_f} 
     \!\! +\frac{N_c}{8\pi^2}\!\sum_{f=u,d,s}\!\! J_{-1}(M_f^2,T,\mu ) 
     + C(T,\mu ),
\end{eqnarray}
where the function $C(T,\mu )$ does not depend on $M$, and therefore cannot be 
determined from the gap equation; it will be found from other arguments in the 
end of this section, but obviously $C(0,0)=0$. The integral 
$J_{-1}(M^2,T,\mu )$ is the immediate generalization of the $T=\mu =0$ case 
(\ref{J_1})
\begin{equation}
\label{J-1}
     J_{-1}(M^2,T,\mu )=-\int_0^{M^2}\!\!\! J_0(M^2,T,\mu )\ud M^2
     =J_{-1}(M^2)+J^{\mathrm{med}}_{-1}(M^2,T,\mu ),
\end{equation}
where the medium contribution to $J_{-1}(M^2,T,\mu )$ is
\begin{eqnarray}
     J_{-1}^{\mathrm{med}}(M^2,T,\mu )&\!\! =\!\!& 4\int_0^{M^2}\!\!\!\ud M^2 
     \int^\infty_0\!\ud |\vec{p}_E||\vec{p}_E|^2 \hat{\rho}_{\Lambda}
     \frac{n_{q}+n_{\overline{q}}}{E_p} =  
     4\int^\infty_0\!\ud |\vec{p}_E||\vec{p}_E|^2\hat{\rho}_{\Lambda\vec{p}_E}
     \int^{M^2}_0\!\!\!\ud M^2\,\frac{n_{q}+n_{\overline{q}}}{E_p}
     \nonumber\\
     &\!\! =\!\!&8\int^\infty_0\!\ud |\vec{p}_E||\vec{p}_E|^2
     \hat{\rho}_{\Lambda\vec{p}_E}\left(2\left(E_p(M)-E_p(0)\right) 
     +T\ln\frac{n_{q M}n_{\overline{q} M}}{n_{q 0}n_{\overline{q} 0}}\right)
     \nonumber\\
     &\!\! =\!\!&8T\int^\infty_0\!\ud |\vec{p}_E||\vec{p}_E|^2
     \hat{\rho}_{\Lambda\vec{p}_E}\ln\frac{\left(1+e^{-\frac{E_p(0)-\mu}{T}}\right)
     \left(1+e^{-\frac{E_p(0)+\mu}{T}}\right)}{\left(1+e^{-\frac{E_p(M)-\mu}{T}}
     \right)\left(1+e^{-\frac{E_p(M)+\mu}{T}}\right)}
\end{eqnarray}
with $n_{q0}, n_{\overline{q}0}$ and $n_{qM}, n_{\overline{q}M}$ referring to the 
occupation numbers for massless and massive particles correspondingly, $E_p(M)
=\sqrt{M^2+\vec{p}^{\, 2}_E}$, $E_p(0)=|\vec{p}_E|$. In spite of the fact that 
the integral $J_{-1}^{\mathrm{med}}(M^2,T,\mu )$ is convergent, we still keep the 
regularization $\hat{\rho}_\Lambda$ to be consistent. Note that the action of 
the operator $\hat{\rho}_\Lambda$ (see eq. (\ref{reg-2})) on any smooth 
function, depending on $M^2$ through the energy, $f(E_p(M))$, can also be 
expressed in terms of momentum as $\hat{\rho}_\Lambda f(E_p)=
\hat{\rho}_{\Lambda\vec{p}_E}f(E_p)$, where
\begin{equation}
     \hat{\rho}_{\Lambda\vec{p}_E}
     =1-\left(1-\Lambda^2 \frac{\partial}{\partial \vec{p}_E^{\, 2}}\right)
     \exp\left(\Lambda^2\frac{\partial}{\partial\vec{p}_E^{\, 2}}\right). 
\end{equation}

Then, noting that, for instance,
\begin{eqnarray}
     &&\int^\infty_0\!\ud |\vec{p}_E||\vec{p}_E|^2 
     \hat{\rho}_{\Lambda\vec{p}_E}\ln\left(1+e^{-\frac{E_p(M)-\mu}{T}}\right)
     \nonumber\\
     &\!\! =\!\!&\hat{\rho}_{\Lambda}\left(\left.\frac{|\vec{p}_E|^3}{3}\ln
     \left(1+e^{-\frac{E_p(M)-\mu}{T}}\right)\right|^{\infty}_0\right) 
     -\hat{\rho}_{\Lambda}\int^\infty_0\!\ud |\vec{p}_E|\frac{|\vec{p}_E|^3}{3}
     \frac{\partial}{\partial |\vec{p}_E|}\ln
     \left(1+e^{-\frac{E_p(M)-\mu}{T}}\right)
     \nonumber\\
     &\!\! =\!\!&\hat{\rho}_\Lambda\int^\infty_0\!\ud|\vec{p}_E|
     \frac{|\vec{p}_E|^4n_{qM}}{3TE_p(M)}=\frac{1}{3T}\int^\infty_0\!\ud 
     |\vec{p}_E||\vec{p}_E|^4\hat{\rho}_{\Lambda\vec{p}_E}\frac{n_{qM}}{E_p(M)}
\end{eqnarray}
where we used the fact the surface term disappears, we get f\mbox{}inally
\begin{equation}
\label{J-1med}
     J_{-1}^{\mathrm{med}}(M^2,T,\mu )=-\frac{8}{3}\int^\infty_0\ud 
     |\vec{p}_E||\vec{p}_E|^4\hat{\rho}_{\Lambda\vec{p}_E}\left(
     \frac{n_{qM}+n_{\overline{q}M}}{E_p(M)}-
     \frac{n_{q0}+n_{\overline{q}0}}{E_p(0)}\right).
\end{equation}

It is important to realize that the expansion of the integral (\ref{J-1med}) 
for small values of $T$ and at $\mu =0$ starts from the term  
\begin{equation}
\label{limit}
     J^{\mathrm{med}}_{-1}(M^2,T,0)=\frac{16}{3}\int_0^\infty
     \frac{|\vec{p}_E|^3\ud |\vec{p}_E|}{1+e^{{\frac{|\vec{p}_E|}{T}}}}
     +{\cal O}(T^{\frac{5}{2}}M^{\frac{3}{2}}e^{-\frac{M}{T}})=\frac{14}{45}
     \pi^4T^4+{\cal O}(T^{\frac{5}{2}}M^{\frac{3}{2}}e^{-\frac{M}{T}}).
\end{equation} 
This leading contribution in $T$ arises from the combination $\sim (n_{q0}+
n_{\overline{q}0})$ and does not depend on the cut-of\mbox{}f, i.e. the 
Pauli-Villars regulator $\hat{\rho}_{\Lambda\vec{p}_E}$ does not af\mbox{}fect 
the leading order of the low-temperature asymptotics of the integral. To make 
this clear let us consider the typical integral in eq. (\ref{J-1med}) 
\begin{equation}
\label{int4}
     \int^\infty_0\ud |\vec{p}_E||\vec{p}_E|^4\frac{n_{qM}}{E_p(M)}
     =\int^\infty_M\frac{|\vec{p}_E|^3\ud E_p}{1+e^{\frac{E_p(M)}{T}}} 
     =T^4\int^\infty_0\!\ud x\frac{(x^2+2x\frac{M}{T})^{\frac{3}{2}}}{1+
     e^{x+\frac{M}{T}}}.
\end{equation}
If $M=0$, then the integral can be evaluated explicitly and is found to be
$\frac{7\pi^4}{120}T^4$, leading us to the result (\ref{limit}). If $M\neq 0$, 
we obtain at once the estimate at small $T$ 
\begin{equation}
     \int^\infty_0\!\ud x\frac{(x^2+2x\frac{M}{T})^{\frac{3}{2}}}{1+
     e^{x+\frac{M}{T}}}
     =\frac{3\sqrt{\pi}}{4}\left(2\frac{M}{T}\right)^{\frac{3}{2}}
     e^{-\frac{M}{T}} 
     \left(1+{\cal O}(T)\right).   
\end{equation}
We conclude that the integral vanishes exponentially for small $T$ and does 
not contribute to the leading order term in eq. (\ref{limit}). It is then clear
that the action of the Pauli-Villars regulator $\hat{\rho}_{\Lambda\vec{p}_E}$ on 
the integrand, which consists in subtracting the contribution of the massive 
Pauli-Villars states, will not af\mbox{}fect the leading term as well. 

One might worry that the low-temperature expansion of the integral 
(\ref{limit}) starts from the unphysical contribution which corresponds to the 
massless quark states. This fear would be valid if the potential had not the 
term $C(T,\mu )$. Let us f\mbox{}ix the $M_f$-independent function $C(T,\mu )$ 
in eq. (\ref{effpot-t}) to avoid the problem. For that it is instructive to 
compare the matter quark-loop part of the thermodynamic potential obtained 
here (i.e. the $J^{\mathrm{med}}_{-1}$-part) with the corresponding result of the 
standard NJL approach. There is only one dif\mbox{}ference between such 
calculations: we use the Pauli-Villars subtractions instead of a 3-dimensional 
cut-of\mbox{}f $\Lambda_3$. Therefore we can expect that if one removes the 
regularizations in both approaches ($\hat{\rho}_{\Lambda\vec{p}_E}\to 1$ and 
$\Lambda_3\to\infty$) the f\mbox{}inite matter part of the thermodynamic
potentials must coincide. From this requirement of consistency we f\mbox{}ind
\begin{equation}
\label{FTP}
       C(T,\mu )= -\frac{N_c}{\pi^2}\int_0^\infty\ud |\vec{p}_E||\vec{p}_E|^4
       \frac{n_{q0}+n_{\bar q0}}{E_p(0)}\ \to\ C(T,0)
       =-\frac{7N_cN_f}{180}\pi^2T^4.
\end{equation}
As a result, the unwanted massless quark contribution to the vacuum energy
at small $T$ disappears.

\subsection{Effective number of quark degrees of freedom}  
A quantity of interest related with the thermodynamic potential is the number 
of quark degrees of freedom present at a certain temperature. For that we 
consider the quark pressure dif\mbox{}ference from the zero-temperature value 
\begin{equation}
\label{nu}
      \nu (T) = \frac{p(T)-p(0)}{\pi^2T^4/90}, 
\end{equation}
in order that the total $\nu (0)=0$. Here $p(T)=-V(M^*_f,T,\mu =0)$, where 
$M^*_f$ denotes the gap equation solution at a given $T$. Dividing the 
pressure in eq. (\ref{nu}) by $\pi^2T^4/90$, the result is presented in 
Stefan-Boltzmann units, i.e. one explicitly counts the number of relevant 
degrees of freedom. 
 
\begin{figure}[t]
\PSfig{0.8}{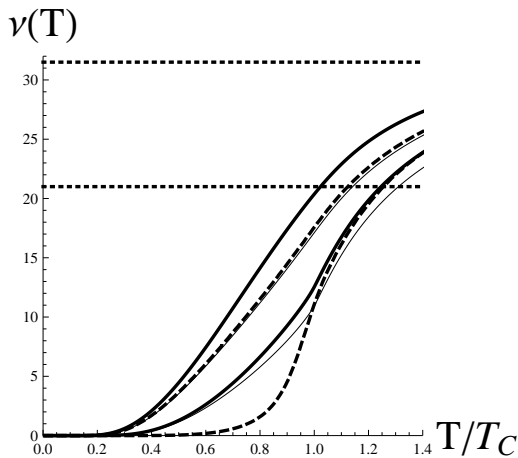}
%\PSfig{1.2}{fig1}
\caption{The number of ef\mbox{}fective degrees of freedom associated with 
   $N_f=3$ f\mbox{}lavor quarks as a function of $T/T_c$ at $\mu =0$ in the 
   NJL model with 't Hooft and eight-quark interactions (bold solid curves 
   correspond to our calculations with a f\mbox{}inite cut-of\mbox{}f, and 
   thin solid curves show the same patterns when one removes the 
   regulator %taking the limit $\Lambda\to\infty$
   in the thermal quark energy integral) 
   compared with the result of Fukushima \cite{Fukushima2:2008} (long dashes). 
   The critical temperature is $T_c=190\,\mbox{MeV}$ (f\mbox{}inite 
   cut-of\mbox{}f), $T_c=179\,\mbox{MeV}$ %($\Lambda =\infty$)
   (no regulator) for parameter 
   set (a) (upper bold solid line, and upper thin solid line, 
   correspondingly), and $T_c=135\,\mbox{MeV}$ (f\mbox{}inite cut-of\mbox{}f), 
   $T_c=132\,\mbox{MeV}$ %($\Lambda =\infty$) 
   (no regulator)for set (b) (lower bold solid 
   line, and lower f\mbox{}ine solid line, correspondingly). The 
   long-dashed-curves are taken from f\mbox{}ig. 3 of paper 
   \cite{Fukushima2:2008}: 
   the upper one corresponds to the NJL model with 't Hooft interactions; 
   the lower one to the NJL model with the Polyakov loop. The short-dashed 
   horizontal lines correspond to the asymptotic high-temperature 
   Stefan-Boltzmann (ideal gas) limit for the theory of massless fermions 
   with three and two f\mbox{}lavors respectively.} 
\label{fig-1}
\end{figure}

It is known that in the case of massless quarks of two and three f\mbox{}lavors
the fermionic degrees of freedom at $T>T_c$ are estimated as $\nu =(7/8)\times 
3\times 2\times 4=21$ and $\nu =(7/8)\times 3\times 3\times 4=31.5$, 
respectively. The system studied here consists of three-types of light quarks: 
$u,d,s$. Hence $\nu$ is expected to be in the interval $21<\nu <31.5$ in the 
region $T>T_c$, where chiral symmetry is ``restored'' (up to the explicit 
symmetry breaking ef\mbox{}fects caused by the current quark masses). Indeed, 
the solid curves $\nu (T/T_c)$, plotted for the sets (a) and (b) in f\mbox{}ig.
\ref{fig-1}, at some values $T/T_c>1$ enter the interval and approach 
asymptotically the upper bound $\nu =31.5$ at high $T$: in particular, we have 
already at $T/T_c =2.5$ that $\nu (2.5)=30.95$ (set (a)) and $\nu (2.5) =30.25$
(set (b)). This is too fast as compared with the lattice estimates 
\cite{Bazavov:2009} in 2+1 f\mbox{}lavor QCD, but the dif\mbox{}ference can be 
ascribed to the simplif\mbox{}ications introduced by the model under 
consideration (the essential dif\mbox{}ference is that the NJL model does not 
possess the quark-conf\mbox{}inement property of QCD).  

We can gain some understanding of the asymptotic behavior of $\nu (T)$ by 
considering eqs. (\ref{J0t}) and (\ref{J-1}). Firstly, it is easy to see that 
the integral $J_0(M^2,T,\mu )\to 0$ at $T\to\infty$. The reason for this is 
very simple and is contained in the integrand $(1-n_q-n_{\bar q})$ which 
vanishes at $T\to\infty$, as it explicitly follows from eqs. (\ref{n-occup}). 
Secondly, eq. (\ref{J-1}) contains $J_0$ as an integrand. Thus we conclude that 
$J_{-1}(M^2,T,\mu )\to 0$ at $T\to\infty$. Next, from the gap equation 
(\ref{gap-t}) it follows that $h_f(T)\to 0$ at $T\to\infty$. Therefore 
$V(M_f,T,\mu )\sim C(T,\mu )$ at large $T$, i.e. we can say that the
asymptotics is totally determined by the term $C(T,\mu )$, yielding $\nu(T\to
\infty )=31.5$, which is independent of any model parameters. This conclusion 
is attractive because it agrees with the general arguments of the previous 
paragraph. In fact, to many readers our conclusion that $\nu (T)$ has the 
correct asymptotics, may seem to be a much more compelling argument for 
f\mbox{}ixing $C(T,\mu )$ than the assumption made that the dif\mbox{}ferent 
cut-of\mbox{}f procedures must give the same result when cut-of\mbox{}fs are 
removed.   

\begin{figure}[h]
\PSfig{0.6}{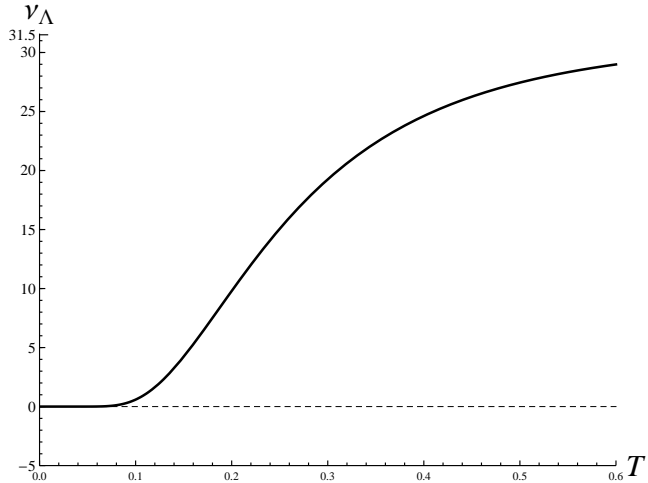}
%\PSfig{0.6}{fig2}
\caption{Solid line: degrees of freedom due to the Pauli-Villars states, 
        $\nu_{\Lambda}(T)$, at zero chemical potential, as function of $T$ 
        [GeV]. Dashed line: the same quantity at $\Lambda\to\infty$.} 
\label{fig-1a}
\end{figure}

A clear insight into the origin of this result can be obtained by considering 
the contribution of $C(T,\mu )$ and the term $\sim (n_{q0}+n_{\bar q0})$ in eq. 
(\ref{J-1med}) to the number of ef\mbox{}fective degrees of freedom $\nu (T)$, 
because exactly these terms determine the correct low-temperature
behavior of the function $\nu (T)$. We designate this contribution by 
$\nu_\Lambda (T)$, and plot it in f\mbox{}ig. \ref{fig-1a}. Thus, we consider 
the following function of temperature 
\begin{equation}
\label{intconst} 
     \nu_\Lambda (T)=\frac{90 N_c}{\pi^4 T^4}\int_0^\infty |\vec{p}_E|^4
     \ud |\vec{p}_E|\,(1-\hat\rho_{\Lambda \vec{p}_E})
     \frac{n_{q0}+n_{\bar q0}}{E_p(0)}
\end{equation} 
calculated at a f\mbox{}ixed value of the cut-of\mbox{}f parameter $\Lambda$ 
(the value taken is the same for both parameter sets considered, see Table I). 
At f\mbox{}irst glance one might wish to associate $\nu_\Lambda (T)$ with the 
contribution of massless states, introduced to $J_{-1}^{\mathrm{med}}$ by 
assigning the low limit $M=0$ to the integral (\ref{J-1}). However, this 
expectation is potentially fallacious. One can easily see from eq. 
(\ref{intconst}) that $\nu_{\Lambda}(T)$ vanishes for all $T$ if the the 
integral is not regularized $\hat\rho_{\Lambda \vec{p}_E}\to 1$. Hence only the 
auxiliary Pauli-Villars states of mass $\Lambda$ contribute to $\nu_\Lambda(T)$ 
\begin{equation}
\label{intconst-2} 
     \nu_\Lambda (T)=\frac{90 N_c}{\pi^4 T^4}\int_0^\infty |\vec{p}_E|^4
     \ud |\vec{p}_E|\left(1-\Lambda^2\frac{\partial}{\partial\vec{p}_E^{\,2}}   
     \right)\frac{n_{q\Lambda}+n_{\bar q\Lambda}}{E_p(\Lambda )}.
\end{equation} 
This contribution is given by an integral with a positive integrand. Here we 
have two dimensionful parameters, $T$ and $\Lambda$. Therefore, at large $T$ 
the series for $\nu_{\Lambda}(T)$ can be organized in powers of the 
dimensionless ratio $\Lambda/T$, i.e.
\begin{equation}
\label{sb-1}
     \nu_\Lambda (T)=\frac{180N_c}{\pi^4}\int_{\frac{\Lambda}{T}}^\infty\!\ud x
     \frac{(x^2-\frac{\Lambda^2}{T^2})^{\frac{3}{2}}}{1+e^x}\left[1+
     \frac{\Lambda^2}{2T^2x^2}\left(1+\frac{xe^x}{1+e^x}\right)\right]=   
     \frac{21N_c}{2}\left[1+
     {\cal O}\left(\frac{\Lambda}{T}\right)\right].
\end{equation} 
A useful observation is that although the heavy mass states determine the value
of the integral, it still has the correct asymptotics at large $T$, which does 
not depend on $\Lambda$. In other words, the integral $\sim (n_{q0}+n_{\bar q0})$
in eq. (\ref{J-1med}) must vanish at $T\to\infty$, in order that the function 
$\nu (T)$ has the right asymptotic behavior. This really happens, due
to the Pauli-Villars regulator $\hat\rho_{\Lambda \vec{p}_E}$. In f\mbox{}ig. 
\ref{fig-1a} one can see that up to around $T\sim 100\,\mbox{MeV}$ the 
contribution to the ef\mbox{}fective number of degrees of freedom is 
practically constant and nearly zero, then it increases and reaches 
asymptotically the value $31.5$. The constant dashed curve shows the same 
quantity if the integral is not regularized. Such a big dif\mbox{}ference 
in the behavior of $\nu_\Lambda (T)$ translates in $\nu (T)$ (see f\mbox{}ig. 
\ref{fig-1}) to an enhancement in the number of quark degrees of freedom 
around the critical temperature in the presence of a regulator. 

%%%%%%%%%%%%%%%%%%%%%%%%%%%%%%%%%    Fig.1     %%%%%%%%%%%%%%%%%%%%%%%%%%%%%%%
%%%%%%%%%%%%%%%%%%%%%%%%%%%%%%%%%%%%%%%%%%%%%%%%%%%%%%%%%%%%%%%%%%%%%%%%%%%%

%%%%%%%%%%%%%%%%%%%%%%%%%%%%%%%    Fig.2    %%%%%%%%%%%%%%%%%%%%%%%%%%%%%%%%
%%%%%%%%%%%%%%%%%%%%%%%%%%%%%%%%%%%%%%%%%%%%%%%%%%%%%%%%%%%%%%%%%%%%%%%%%%%%
%\vspace{0.5cm}

The conclusion that the Pauli-Villars regularization leads to the correct 
asymptotics is attractive because in the NJL model with 3-dimensional 
cut-of\mbox{}f the Stefan-Boltzmann limit can be reached only when the 
cut-of\mbox{}f is removed in the matter integrals \cite{Klevansky:1994}, which 
are f\mbox{}inite in themselves. Such a selective removal of the cut-of\mbox{}f 
must be taken ``cum grano salis'', since technically the NJL model requires 
the presence of a f\mbox{}inite ultraviolet cut-of\mbox{}f throughout all 
integrals. Nevertheless, one might think of criticizing our result on the 
grounds that we seem to get the correct asymptotics due to the auxiliary and 
therefore unphysical Pauli-Villars terms. This is certainly not true. To see 
this let us return to eq. (\ref{J-1med}) and consider now the contribution of 
the term $\sim (n_{qM}+n_{\bar qM})$ to $\nu(T)$. (Our arguments will be based on 
considering the simplif\mbox{}ied case $M_u^*=M_d^*=M_s^*=M_*$. We need not be 
rigorous in the following discussion, because the unitary symmetry breaking 
ef\mbox{}fects are unimportant for the asymptotics.) The integral describes the 
thermal energy of massive quarks, but does not contribute at large $T$
\begin{equation}
\label{sb-2}
     \nu_{M_*} (T) =\frac{90 N_c}{\pi^4 T^4}\int_0^\infty |\vec{p}_E|^4\ud 
     |\vec{p}_E|\hat\rho_{\Lambda \vec{p}_E}\frac{n_{qM_*}+n_{\bar qM_*}}{E_p(M_*)} 
     =\frac{21N_c}{2}\left[1-1+{\cal O}
     \left(\frac{M_*}{T},\frac{\sqrt{\Lambda^2+M_*^2}}{T}\right)\right].
\end{equation} 
The vanishing result is a consequence of a total cancellation of two
contributions: the f\mbox{}irst, $1$, in the square brackets represents the 
contribution of the physical states with mass $M_*$; the second, $-1$, comes 
from the Pauli-Villars regulator. What is interesting here is that the second 
term, if one joins it with the other unphysical contributions of eq. 
(\ref{sb-1}), cancels them entirely. In other words, the correct large $T$ 
asymptotics in (\ref{nu}) can equally be assigned to the pure physical states 
contribution as well.  
          
Let us compare our results with the recent estimates of Fukushima 
\cite{Fukushima2:2008}, made on the basis of the three-f\mbox{}lavor NJL model 
with and without the Polyakov loop. The starting values are very similar: our 
curve for the set (a) agrees well with the Fukushima estimate made in the 
standard NJL model approach with the 't Hooft six-quark interactions. Indeed, 
if we remove the cut-of\mbox{}f dependence in the quark number occupation 
integrals (\ref{J-1med}), like it has been done in \cite{Fukushima2:2008}, the 
curves almost coincide: compare the upper thin solid line of our set (a), 
where the regulator has been removed, with the long dashed curve of Fukushima,
almost on top of it in f\mbox{}ig. \ref{fig-1}. Taking systematically into 
account the f\mbox{}inite value of the cut-of\mbox{}f, we obtain the upper 
bold curve corresponding to the set (a). The set (b), compared with set (a), 
shows a rather strong (more than 50\%) suppression of the abundant 
artif\mbox{}icial quark excitations at $T/T_c>0.2$ due to the large 
OZI-violating eight-quark interactions (see lower bold curve in f\mbox{}ig. 
\ref{fig-1} for the f\mbox{}inite cut-of\mbox{}f result and lower thin solid 
line, where the cut-of\mbox{}f condition has been relaxed. Notice that in both
cases (a) and (b) the f\mbox{}inite cut-of\mbox{}f leads to larger values of 
$\nu$, because $\nu_\Lambda (T)$ is non-zero and positive at $\Lambda =\infty$; 
this ef\mbox{}fect is more pronounced in the case of set (a)). Although with 
set (b) the model still fails to describe accurately the pressure at 
$0.3<T/T_c<1$, it leads to an improved description of the number of quark 
degrees of freedom as compared to set (a). Thus the OZI-violating interactions
could potentially play an important role in the description of quark 
excitations. At least one should not exclude the set (b) if one includes the 
Polyakov loop ef\mbox{}fects in the framework of ef\mbox{}fective NJL-type 
models with multi-quark interactions, see \cite{Bhattacharyya:2010} for 
results with the 3-dimensional cut-of\mbox{}f.

\section{Phase diagram}
\begin{figure}[t]
\PSfig{0.6}{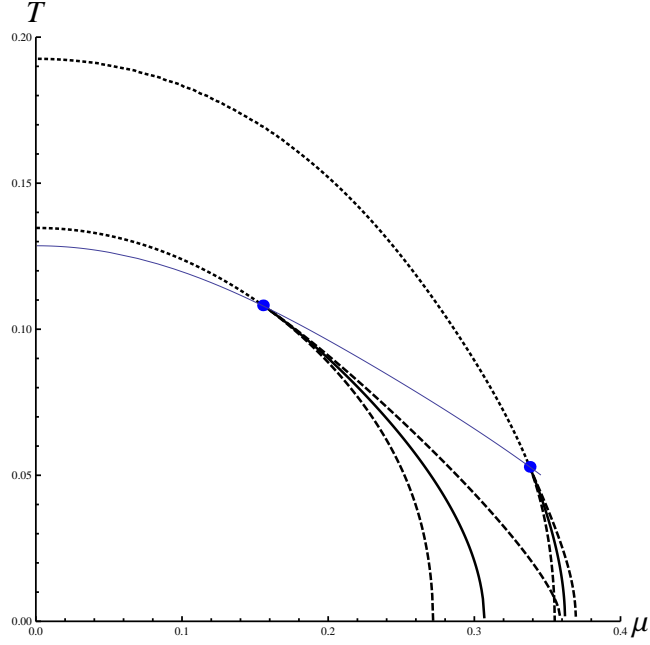}
%\PSfig{0.6}{fig3}
\caption{Phase diagrams indicating f\mbox{}irst order phase transition lines: 
         the short solid line corresponds to parameter set (a) of weak $8q$ 
         coupling constant; the long solid line is for set (b) with strong $8q$ 
         coupling. Dashed lines are spinodals. Circles (blue online) indicate 
         the critical endpoints (CEP) for sets (a) and (b) respectively and 
         dotted lines correspond to the crossover region. The thin solid line 
         (blue online) represent all CEP obtained by varying the $8q$ coupling 
         $g_1$ and keeping as before the meson mass spectra at $T=\mu=0$ 
         unchanged, except for the $\sigma$-meson mass. All units are in GeV.} 
\label{fig-2}
\end{figure}

Fig. \ref{fig-2} shows the phase diagrams for the two sets of parameters. One 
observes a larger window for the f\mbox{}irst order transition regime in the 
case of stronger $8q$ interaction coupling $g_1$, the critical endpoint is 
situated at $(\mu_E,T_E)=(155,108)$ MeV, whereas for the smaller coupling it 
is at $(\mu_E,T_E)=(338,53)$ MeV. The temperature at zero chemical potential, 
in the crossover regime, is substantially smaller for the large $8q$ coupling 
case, around $T_c\simeq 135\,\mbox{MeV}$ compared to $T_c\simeq 193\,
\mbox{MeV}$ for the sets (b) and (a) respectively \cite{Osipov:2007b}. The 
approximate values for $T_c$ were obtained as usual, through the condition 
${\rm d}^2M/{\rm d}T^2=0$ of strongest change in the slope of the constituent 
quark masses.

%%%%%%%%%%%%%%%%%%%%%%%%      Fig.3     %%%%%%%%%%%%%%%%%%%%%%%%%%%%%%%%%%%%%

%%%%%%%%%%%%%%%%%%%%%%%%%%%%%%%%%%%%%%%%%%%%%%%%%%%%%%%%%%%%%%%%%%%%%%%%%%%%%
\vspace{0.5cm}

%%%%%%%%%%%%%%%%%%%%%%%%       Fig.4      %%%%%%%%%%%%%%%%%%%%%%%%%%%%%%%%%%%
\begin{figure}[h]
\PSfig{0.5}{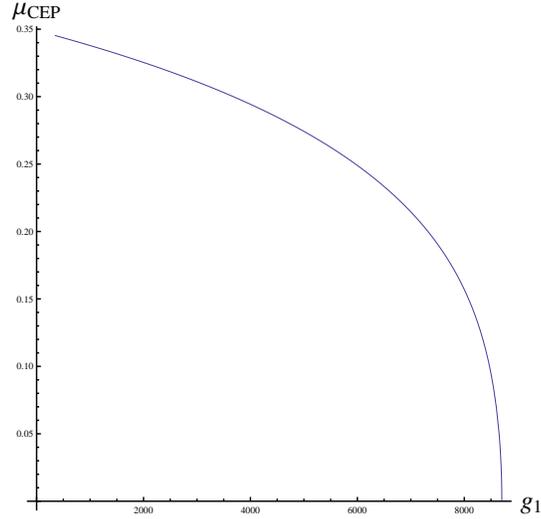}
%\PSfig{0.6}{fig4}
\caption{The line of CEP described in f\mbox{}ig. \ref{fig-2}, now in the 
         $\mu_{CEP}, g_1$ plane.}
\label{fig-2a}
\end{figure}
%%%%%%%%%%%%%%%%%%%%%%%%%%%%%%%%%%%%%%%%%%%%%%%%%%%%%%%%%%%%%%%%%%%%%%%%%%%%%%

The length of the line of f\mbox{}irst order transitions can be increased 
further with increase of the $8q$ coupling $g_1$ and after some critical value 
it goes all the way through to the $\mu =0$ end. A corresponding set of 
parameters has been considered in \cite{Osipov:2008}. In this way we obtain the
thin solid line in f\mbox{}igs. \ref{fig-2}, \ref{fig-2a} that connects all CEP
obtained by varying $g_1$ in the allowed interval for stability of the 
ef\mbox{}fective potential. 

However we point out that also in the case of strong $8q$ couplings our results
have with other approaches in common that the dynamical $u$ and $d$ quark 
masses suf\mbox{}fer a considerable reduction at the transition temperatures, 
while the strange quark mass remains roughly the same. For instance for set (b)
at $\mu=170\,\mbox{MeV}$ at the f\mbox{}irst order transition temperature 
$T=103\,\mbox{MeV}$, $M_u$ jumps from $M_u=208\,\mbox{MeV}$ to $128\,
\mbox{MeV}$, while the strange quark mass changes only from $M_s=487\,
\mbox{MeV}$ to $448\,\mbox{MeV}$. By increasing the temperature keeping $\mu$ 
f\mbox{}ixed, one obtains at $T=160\,\mbox{MeV}$ the values $M_u=24\,
\mbox{MeV}$ and $M_s=381\,\mbox{MeV}$. Prof\mbox{}iles of the quark masses are 
given in \cite{Moreira:2010} for the $T=0,\mu\ne 0$ case. The slow decrease of 
the strange quark mass is present in studies involving realistic values for the
strange current quark mass \cite{Fukushima:2008}. 

The position of the critical endpoint in the three f\mbox{}lavor NJL model with
$U_A(1)$ breaking has been analyzed in \cite{Fukushima2:2008}, with the 
parameter set of \cite{Hatsuda:1994}, $(\mu_E,T_E)=(324,48)\,\mbox{MeV}$ in 
comparison with the case with inclusion of the Polyakov loop, $(\mu_E,T_E)=
(313,102)\,\mbox{MeV}$. The increase of $T_E$ is explained by the suppression 
of the artif\mbox{}icial quark excitations at f\mbox{}inite temperature and 
density in the presence of the Polyakov loop contribution 
\cite{Fukushima2:2008}. Our set (a) with $(\mu_E,T_E)=(338,53)\,\mbox{MeV}$ is 
in reasonable agreement with the above estimates.

Our result for set (b), $(\mu_E,T_E)=(155,108)\,\mbox{MeV}$, yields a twice 
lower value for the critical $\mu_E$ as compared to set (a). This new feature 
is related with the OZI-violating eight-quark interactions: it is well-known 
that NJL-type models without $8q$-forces have the tendency in common to lead 
to a critical endpoint at relatively high density, above $\mu_E\sim 300\,
\mbox{MeV}$ \cite{Fukushima2:2008}. The small value found here is a good 
indicator that the main force responsible for dynamical chiral symmetry 
breaking has changed, being now associated with the 't Hooft $6q$-interactions.

Let us also take notice of the dif\mbox{}ference between the critical 
temperatures for the case (b), $T_c=108\,\mbox{MeV}$, and the Fukushima's case 
with Polyakov loop, $T_c=204.8\,\mbox{MeV}$. We expect that the inclusion of 
the Polyakov loop in our analysis will increase somewhat the value of $T_c$.  

The NJL model with $8q$ interactions and without $U_A(1)$ symmetry breaking 
has been analyzed as well for the $SU(2)$ f\mbox{}lavor case 
\cite{Kashiwa:2006,Kashiwa:2008}. Although the $8q$ forces are not needed to 
stabilize the ef\mbox{}fective potential of the model in the two f\mbox{}lavor 
case, the same tendency as above in i) was observed, for instance: 
$(\mu_E,T_E)=(276,62)\,\mbox{MeV}$ with the $8q$ interactions, whereas 
$(\mu_E,T_E)=(330,47)\,\mbox{MeV}$ without them \cite{Kashiwa:2008}.

Another variant without $8q$ terms of the two f\mbox{}lavor NJL model, 
including the vector-isoscalar interactions, which induce an ef\mbox{}fective 
chemical potential, has been considered a long time ago \cite{Asakawa}. In 
this case, the critical endpoint is located at larger $\mu$ and lower 
temperatures $(\mu_E,T_E)=(350,40)\,\mbox{MeV}$. This ef\mbox{}fect has been 
studied also recently in \cite{Kashiwa:2008,Fukushima2:2008}. 
   
Further results concerning the position of the critical endpoint within other 
model approaches is given in the review \cite{Stephanov:2004} (note the 
notation there is in terms of $\mu_B= 3\mu$), in comparison with lattice 
results and the freezeout points extracted from heavy-ion experiments. For our 
two sets the critical endpoints are situated in case (a) slightly above and in 
case (b) slightly below the freezeout points obtained at dif\mbox{}ferent 
collision energies \cite{Braun-Munzinger:2003,Stephanov:2004}.

\section{Conclusions}
The thermodynamic potential of the three f\mbox{}lavor NJL model with 't Hooft 
and $8q$ interactions has been obtained in stationary phase approximation (at 
leading order) and using the Pauli-Villars regularization in quark loops (at 
one-loop level). 

The main conclusions of this work can be classif\mbox{}ied as follows. 

Firstly, we argued that a non-renormalizable ef\mbox{}fective theory at 
f\mbox{}inite temperature and chemical potential, can be self-consistently 
studied with the use of the Pauli-Villars regularization. The importance 
of the regulator in the matter parts for consistency has also been discussed 
in \cite{Florkowski:1997} in connection with correlators. Instead, our 
present study is devoted to the construction of the thermodinamic potential. 
Unlike the conventional approach with 3-dimensional cut-of\mbox{}f, the 
Pauli-Villars technique leads to the right asymptotic behavior of relevant 
thermodynamic observables. It is one of the main results of this paper. 

Secondly, we quantif\mbox{}ied the ef\mbox{}fect of the new $8q$ terms on the 
number of degrees of freedom and on the phase diagram. We observe that in the 
large $8q$ coupling regime a strong depletion of the number of degrees of 
freedom below $T_c$ is reached in comparison with the weak coupling case, 
working in the same direction as the ef\mbox{}fect produced through inclusion 
of the Polyakov loop in the model without $8q$ interactions. 

Thirdly, we conclude that the NJL model with the $8q$ stabilizing interactions 
does not impede the possibility of having a phase diagram consisting only of 
f\mbox{}irst order transitions even for realistic quark masses. This will 
depend on the strength of the OZI-violating $8q$ interactions. At $\mu =0$ 
there is growing evidence from lattice calculations that the transition is a 
crossover \cite{Aoki:2006}. This would set an upper limit for the $8q$ 
coupling, which nevertheless can be suf\mbox{}f\mbox{}iciently strong to 
trigger the f\mbox{}irst order transitions regime at low values of $\mu\ne 0$. 
This point deserves to be studied more carefully especially because it can 
help us to clarify the quark dynamics which is responsible for the mechanism 
of spontaneous chiral symmetry breaking. Indeed as it is shown above the shift 
of the critical endpoint to lower values of $\mu_E$ is possible only when the 
$6q$ interactions are responsible for the chiral phase transition (set (b)); 
however if the $4q$ coupling $G$ exceeds its critical value, the $4q$ 
interactions drive the chiral phase transition in NJL-type models and as a 
result the critical endpoint is located at large values of $\mu_E$ (set (a)). 
  
\vspace{0.5cm}
    
\centerline{\bf ACKNOWLEDGMENTS}
\vspace{0.5cm}

This work has been supported in part by grants of Funda\ca o para a 
Ci\^encia e Tecnologia, FEDER, OE, POCI 2010,  CERN/FP/83510/2008, 
SFRH/BD/13528/2003 and Centro de F\ii sica Computacional, unit 405. 

We acknowledge the support of the European Community-Research 
Infrastructure Integrating Activity "Study of Strongly Interacting Matter" 
(acronym HadronPhysics2, Grant Agreement n. 227431) under the Seventh 
Framework Programme of EU.

\end{document}